# Evolution of the Supercluster-Void Network

Patrick Frisch[1], Jaan Einasto[2], Maret Einasto[2], Wolfram Freudling[3,4], Klaus J. Fricke[1], Mirt Gramann[2,5], Veikko Saar[2], Ott Toomet[2]

[1] University Observatory, D-37083 Göttingen
[2] Tartu Astrophysical Observatory, EE-2444 Tõravere, Estonia
[3] Space Telescope – European Coordinating Facility, D-85748 Garching
[4] European Southern Observatory, D-85748 Garching
[5] Princeton University Observatory, Peyton Hall, Princeton, NJ 08544

**Abstract.** Recently, the observed cellular nature of the large-scale structure of the Universe with its quasi-regular pattern of superclusters and voids has been pointed out by several authors. In this paper, we investigate properties of the initial power spectrum which lead to prediction of structure consistent with these observations. For this purpose, we analyze the evolution of structure within four sets of 2- and 3-dimensional cosmological models, which differ in their initial power spectrum. The models include HDM and CDM models as well as double power-law models. We discuss in detail the impact of model parameters such as the large scale and small scale power and the position and height of the maxima of the power spectra on the predicted structure. Several statistical techniques were employed to compare the models with observations. They include the analysis of the distribution of voids defined by rich and poor clusters of galaxies, voids defined by galaxies, clusters and superclusters. In addition, the cluster correlation function is compared.

We conclude that the observed regular distribution of superclusters and voids can be reproduced only if the spectrum of density fluctuations has a well-defined maximum. The wavelength of the maximum determines the scale of the structure. Small-scale fluctuations determine the fine structure of the Universe. Large-scale fluctuations modulate the fine structure and determine the quasi-regular structure on supercluster scales.

The best agreement with observations was observed in the model with the Harrison-Zeldovich spectrum on large scales, a power index $n \approx -1.5$ on small scales, and a maximum of the power spectrum at $\approx 130\ h^{-1}$ Mpc. In this model the distribution of masses of clusters and superclusters, the correlation function of clusters, and the void distribution reproduce well the respective observed distributions.

In models with no power on large scales all superclusters are equal in mean density, while in models with negative power index on large scales the mass distribution function of clusters is too shallow. In the HDM model (no power on small scales) the cluster-defined voids are completely empty. CDM-models have no well-defined maximum of the spectrum, and the cellular distribution of superclusters and voids is insufficiently developed in this case.

We also investigated the dynamical evolution of the supercluster-void structure. The results show that the basic supercluster-void network is formed very early and is essentially given by initial conditions.

## 1. INTRODUCTION

Several recent studies of the large-scale distribution of clusters of galaxies have found a quasi-regular cellular pattern of superclusters and cluster-defined voids (supervoids) (Broadhurst *et al.* 1990, Bahcall 1991, Tully *et al.* 1992, Einasto *et al.* 1994b, hereafter EETDA, Vettolani *et al.* 1994a,b). Superclusters separate supervoids at fairly regular intervals. The characteristic scale of the supercluster-void network is $\approx 130\ h^{-1}$ Mpc. The regular distribution of superclusters of galaxies can be seen in Figures 5 and 7 by EETDA and Figure 1a of Lindner *et al.* (1994), where rich clusters are plotted in supergalactic rectangular coordinates. An essential property of the distribution is the regularity of the structure: near the supergalactic equatorial plane $z = 0$ there are at fairly regular intervals six rich superclusters. The distribution of rich superclusters in the vertical plane is very similar, see Tully *et al.* (1992), Bahcall (1991) and EETDA. As emphasized by Tully *et al.* (1992) this distribution looks like a three-dimensional chessboard.

Superclusters are made up of a net of galaxy filaments and knots (clusters of galaxies). The same constituents are also found in supervoids which contain filamentary galaxy systems. The large-scale distribution of galaxies in the direction of the Coma and Hercules superclusters was studied by Lindner et al. (1994). In front of these superclusters there is a large low-density region without any rich clusters of galaxies – the Northern Local supervoid. This low-density region embodies numerous galaxy filaments. The basic difference between the constituents of superclusters and supervoids lies in their richness: in superclusters systems of galaxies are much richer and denser than in supervoids.

These observations raise several important questions: What causes the formation of a regular network of superclusters and supervoids? Why do both superclusters and supervoids consist of filamentary structures of different richness?

A large body of numerical simulations of the formation of the structure in the Universe exists in the literature. Models with CDM perturbation spectra have been used (Efstathiou et al. 1985, West et al. 1991, Little and Weinberg 1994), as well as simple power-law models (Efstathiou et al. 1988, Weinberg and Gunn 1990, Beacom et al. 1991, Melott & Shandarin 1993) and combined CDM and HDM models (Davis, Summers and Schleger 1992, Holtzman and Primack 1993, Klypin et al. 1993). The formation of a regular network of superclusters and supervoids has been addressed so far only by Einasto and Gramann (1993). They considered the problem, however, only qualitatively. A quantitative study of the influence of perturbations of various scales using a broad set of initial conditions is still lacking.

The goal of this paper is to model two observed features of the large scale structure. First we focus on the question what kind of initial conditions are required to reproduce the quasi-regular distribution of superclusters and supervoids. Then we study which conditions are needed to explain the filamentary structure of galaxies in superclusters and supervoids. Since initial conditions are given by the power spectrum, our task is to find a suitable initial power spectrum. Direct determinations of the power spectrum on the scales of interest have too large random errors – these scales are close to the size of the largest volume for which data are available. Thus we use an indirect method to investigate the behavior of the Universe on large scales by modelling the evolution of the structure and by comparing results with observed properties using various statistics. For that purpose, we carried out N-body simulation to simulate the distribution of matter. Since the aim is the study of the influence of large-scale modes of density perturbations, we consider a computational box which is considerably larger than the scale of the maximum of the power spectrum.

The paper is organized as follows. A description of the model simulations, and rules to identify the clustered matter associated with galaxies, as well as individual clusters and superclusters is given in §2. In §3 we discuss the statistical tests to be applied. Next we analyze models for various model parameters. We change separately in §4 the position of the maximum; in §5 small-scale power index; in §6 large-scale power index; and in §7 the strength of the maximum. In this analysis of models we use different statistics (sizes of voids defined by different objects, the mass distribution of clusters and superclusters, the correlation function). In §8 we present results of the study of the evolution of the structure in different models. The discussion of our results is given in §9. Finally we summarize principal results of the study.

In this paper $h$ denotes the Hubble constant in units of 100 kms$^{-1}$Mpc$^{-1}$.

## 2. MODELS

### 2.1. Model simulations

We use the particle-mesh code by Gramann (1988) to simulate the evolution of the distribution of mass. Most of the investigations are done on two-dimensional simulations. Two dimensional modelling is not only less demanding in using computer resources, it also facilitates the easy interpretation and graphical representation of results. However, some aspects of the structure can be checked only by three-dimensional modelling, we therefore also use a limited number of three-dimensional models. Principal parameters of models are given in Table 1.

**Table 1.** Data on models used

| Model | D | $\Omega$ | $n_3$ | $m_3$ | $k_t$ | $L$ $h^{-1}$ Mpc | $\sigma_8$ |
|---|---|---|---|---|---|---|---|
| M2pk2 | 2 | 1 | 1 | $-1.5$ | 2 | 256 | 0.69 |
| M2pk4 | 2 | 1 | 1 | $-1.5$ | 4 | 512 | 0.68 |
| M2pk8 | 2 | 1 | 1 | $-1.5$ | 8 | 1024 | 0.68 |
| M2pi1 | 2 | 1 | 1 | $-1$ | 4 | 512 | 0.50 |
| M2pi2 | 2 | 1 | 1 | $-2$ | 4 | 512 | 0.66 |
| M2pi3 | 2 | 1 | 1 | $-3$ | 4 | 512 | 0.75 |
| HDM | 2 | 1 | 1 | $-\infty$ | 4 | 512 | 0.92 |
| M1p | 2 | 1 | $-1.5$ | $-1.5$ | 4 | 512 | 0.67 |
| M1pt | 2 | 1 | $+\infty$ | $-1.5$ | 4 | 512 | 0.66 |
| M2p | 2 | 1 | 1 | $-1.5$ | 4 | 512 | 0.64 |
| N2p | 3 | 1 | 1 | $-1.5$ | 4 | 512 | 0.90 |
| CDM1 | 3 | 1 | 1 | $-0.5$ | 4 | 512 | 0.70 |
| CDM2 | 3 | 0.2 | 1 | $-1.5$ | 4 | 512 | 0.85 |

Designations in the table are as follows: $D$ is the dimension of the simulation; $\Omega$ is the density parameter; $n_3$ and $m_3$ are effective power indices of the spectrum on large and short wavelengths, respectively (index $_3$ indicates that the spectral index corresponds to 3-dimensional case, the index in 2-dimensional case used in 2-d calculations, is $n_2 = n_3 + 1$ and $m_2 = m_3 + 1$); $k_t$ is the wavenumber corresponding to the maximum of the spectrum, wavenumber $k = 1$ corresponds to the size of the computational box $L$; $\sigma_8$ is the dispersion of mass density fluctuations on $8\ h^{-1}$ Mpc scale.

In most cases we use on large scales the Harrison-Zel'dovich spectrum with index $n_3 = 1$. We assume, (see §9.1.) that the maximum of the spectrum lies at wavelength $\lambda_t = 128\ h^{-1}$ Mpc. This wavelength is our scaling parameter; i.e. the length scale of the models is defined by the maximum in the measured power spectrum.

Two-dimensional models were calculated using $512^2$ cells and particles. To investigate the influence of spatial resolution part of the calculations were carried twice using $256^2$ cells and particles. The spatial resolution of these models is $1\ h^{-1}$ Mpc and $2\ h^{-1}$ Mpc, respectively. In 3-dimensional models we use $128^3$ cells and particles, the spatial resolution is $4\ h^{-1}$ Mpc. Our experiments with 2-dimensional models have shown that this resolution is sufficient to isolate clusters and superclusters of galaxies.

To investigate the influence of different parameters of the spectrum in detail we have performed four series of simulations.

In the first series we varied the *power on large scales*. The first model has a simple power law model with density spectrum

$$P(k) = A\kappa^m. \tag{1}$$

Here $\kappa = k/k_t$, $k = L/\lambda$ is the dimensionless wavenumber, $\lambda$ is the wavelength, $k_t = L/\lambda_t$ is a normalizing wavenumber, $A$ is the amplitude, $m$ is the power index. The wavenumber $k$ is taken equal to unity for the whole box length $L$. This one power law model is designated as M1p.

In the second model, designated as M1pt, we use a truncated power law

$$P(k) = \begin{cases} A\kappa^m, & \text{if } k \geq k_t \\ 0, & \text{if } k < k_t. \end{cases} \tag{2}$$

In this model $\kappa = k/k_t$, $k_t = L/\lambda_t$ is the truncation wavenumber, and $\lambda_t$ is the corresponding wavelength.

The third model of the first series is a double power law model with a smooth transition, designated as M2p:

$$P(k) = A\frac{\kappa^n}{1 + \kappa^{n-m}}, \tag{3}$$

If $n - m > 0$, then indices $n$ and $m$ characterize the spectrum at small and large $k$, respectively: at small wavenumbers $\kappa^{n-m} \ll 1$ and we get a simple power law with index $n$, at large wavenumbers we can ignore 1 in the denominator and have a simple power law with index $m$.

In the second series of models we vary the *power index on small scales*, $m$, only. The respective models are designated as M2p1, M2p2, M2p3, the last number corresponds to the absolute value of the spectral index in 3-dimensional case (actual simulations have been performed in 2-D, and the index is respectively lower in absolute value). Instead of a smooth transition between large and small wavelengths in (3) we use a sharp one:

$$P(k) = \begin{cases} A\kappa^m, & \text{if } k \geq k_t \\ A\kappa^n, & \text{if } k < k_t. \end{cases} \tag{4}$$

The HDM model also belongs to this series as the extreme case having no power on small wavelengths. We approximate the spectrum with Harrison-Zeldovich index $n_3 = 1$ on large scales, and accept a sharp truncation at wavenumber $k_t$.

In the third series of models we vary the *scale of the maximum of the power spectrum*, $k_t = 2, 4, 8$. Respective models are designated as M2pk2, M2pk4, M2pk8. We use again a sharp transition between short and large wavenumbers (4). These models correspond to different sizes of boxes, $L$, since we assume that the scale of the maximum of the spectrum, $\lambda_t$, is fixed.

The last series of models is performed to investigate the influence of the *height of the maximum of the power spectrum*. The strong maximum case corresponds to a sharp transition between the two power laws (4); the weak maximum case was taken according to a standard CDM spectrum. We use the spectrum in the form given by Gramann (1988)

$$P(k) = A\frac{k}{\left(1 + \frac{(ak)^2}{\log(1+bk)}\right)^2}. \tag{5}$$

In this formula the parameters are as follows: $a = 4.0/(\Omega h^2)$, $b = 2.4/(\Omega h^2)$, $h$ is the Hubble constant, $\Omega$ is the density parameter in units of the critical closure density, and the wavenumber is dimensional, $k = 2\pi/L$. The values of constants $a$ and $b$ correspond to cube size $L$, expressed in $h^{-1}$ Mpc. In calculations we have adopted the Hubble constant $h = 0.5$.

If $\Omega = 0.9$ then the maximum of the spectrum is located at the wavelength $\lambda_t = 128\ h^{-1}$ Mpc. The maximum of the spectrum is very shallow, therefore we do 3-dimensional simulations only, using two values of the density parameter $\Omega$, models CDM1 and CDM2. The power index in 2-dimensional case differs by one unit from the 3-dimensional case, and for a shallow maximum it shifts to a different wavelength, thus the interpretation of 2-dimensional models in respect of the maximum wavelength is difficult. For the low-density model CDM2 we assume the presence of a cosmological term $\Omega_\Lambda$, where $\Omega + \Omega_\Lambda = 1$. For comparison with 2-dimensional simulations we calculate also a 3-dimensional two-power law model, N2p, using the spectrum (4).

The initial spectrum was generated in all model families using identical particle distributions, thus models of the same family correspond to the same realization of the structure. In this case the differences in the structure depend only on the parameter which is changed in the particular family.

**Fig. 1.** Spectra (at the present epoch) for all models considered. Observed mass spectrum is shown by dots (Gramann and Einasto 1992, Einasto, Gramann and Saar 1993). The wavenumber $k = 1$ corresponds to the size of the computational box. The observed galaxy spectrum is reduced to the mass spectrum. It lies below the model spectra by a factor of 3, due to overcorrection for the bias factor in these papers.

On small scales initial fluctuations were generated by a gaussian random process. In order to minimize the irregularities caused by the small number of large-scale modes, for large scales the initial amplitude of a mode was fixed by the value of the spectrum according to the analytic expression (1) – (5), while the sign of the amplitude was taken at random.

We are also interested in *the evolution of the structure*. For this purpose we identify four epochs, characterized by rms dispersion of spatial density fluctuations on the cell scale (1 $h^{-1}$ Mpc), 0.5, 1, 2, 4. Einasto *et al.* (1994a) derived the dispersion on the scale 1.2 $h^{-1}$ Mpc for a sample in the Virgo supercluster. After correction to the matter density fluctuation they found a value $\sigma = 4$ for the present epoch. This dispersion is characteristic for density smoothing on scales $\approx 1$ $h^{-1}$ Mpc. If the density is smoothed on 2 $h^{-1}$ Mpc scale then the present epoch corresponds to a dispersion $\sigma \approx 3.5$. The dispersion grows almost linearly with the expansion factor $a$, thus the first epoch corresponds approximately to a redshift $1 + z = 8$. From the density power spectrum we calculate also the dispersion on an 8 $h^{-1}$ Mpc scale for the last epoch, $\sigma_8$.

Spectra for all models for the present epoch are plotted in Figure 1. As noted above, the present epoch was identified in 2-dimensional models by the density fluctuations on the 1 $h^{-1}$ Mpc scale, $\sigma_1 = 4$, in 3-d models by fluctuations on the 4 $h^{-1}$ Mpc scale, $\sigma_4 = 2.5$. Fluctuations on 8 $h^{-1}$ Mpc scale, $\sigma_8$, were calculated from the spectrum for the last epoch.

*2.2. Identification of galaxies*

The resulting distribution of particles in the model is interpreted as the distribution of matter in the Universe, which includes both luminous and dark matter. In order to identify galaxies, we assume that galaxy formation occurs only in regions where the density exceeds a certain threshold. Both observations (Einasto *et al.* 1994a) and numerical simulations of galaxy formation (Cen and Ostriker 1992) suggest that the threshold density for galaxies to form is close to the mean density of matter if smoothed on the scale of typical diameters of galaxy systems, $R_g \approx 1\ h^{-1}$ Mpc. In regions of lower density there are no galaxies.

In addition, we assume that in regions with density higher than the mean density galaxies follow the matter distribution. Available observational data support this assumption (Vennik 1986, Hughes 1989). Under these assumptions all test particles in these regions can be considered as galaxies. Depending on the size of the computational box and the number of particles in a particular simulation every particle corresponds to a galaxy of certain mass including the dark corona which contains most of the mass of the galaxy. In 2-dimensional models with $k_t = 4$ there is one particle in a cell of size $1\ h^{-1}$ Mpc, and such a particle corresponds to a mass $0.28 \times 10^{12}\ M_\odot$ (a dwarf galaxy). In 3-dimensional models with $k_t = 4$ there is one particle in a cell of size $4\ h^{-1}$ Mpc which corresponds to a mass $17.8 \times 10^{12}\ M_\odot$ (a massive giant galaxy).

The smoothing length $R_g$ is an additional parameter to the model. Einasto *et al.* (1994a) have demonstrated that the exact choice of this parameter has little influence on the comparison between different models. In addition, trial calculations with models of lower resolution ($R_g = 2\ h^{-1}$ Mpc) have shown that properties of models with either resolution are virtually identical.

### 2.3. Identification of clusters of galaxies

Clusters of galaxies are identified as peaks in the density field. In calculating the density field we use gaussian smoothing with dispersion $R_g = 1\ h^{-1}$ Mpc. Cluster properties are determined by fitting an ellipsoid to the density field at the maximum. The center of the ellipsoid gives the position of the cluster and the volume above a certain threshold density of the ellipsoid – the cluster mass. This threshold density is chosen so that the cluster has dimensions comparable to real clusters of galaxies. Using a trial and error procedure we found that a density threshold in units of the mean density $D_t = 2$ efficiently separates clusters of galaxies from field galaxies. If the threshold density is too low then the resulting clusters are too large in volume and have an elongated shape, this threshold corresponds already to a galaxy filament. The peak density in clusters of galaxies is several hundred in mean density units and the rise of the density near the peak is very rapid. Thus the resulting cluster catalogue is not sensitive to the exact level of the threshold density.

We distinguish between poor and rich clusters of galaxies, depending on the mass of the density enhancement. The transition density between poor and rich clusters is chosen so that their spatial density agrees with those observed for Abell clusters of galaxies, $13.5 \times 10^{-6} h^3$ Mpc$^{-3}$ (Bahcall and Cen 1993). The lower mass limit of poor clusters is chosen in such a way that the spatial density of poor clusters is four times higher than the density of Abell clusters. The actual density of nearby Zwicky clusters is 5.6 times higher than the density of Abell clusters, but since not all of our models contain such a high number of clusters, we use this number appropriate for Zwicky clusters.

### 2.4. Identification of superclusters

Superclusters of galaxies were identified from the density after gaussian smoothing. Superclusters were found as high-density regions in the density field with a smoothing length $R_b$. This smoothing length was chosen considerably larger than the smoothing length $R_g = 1\ h^{-1}$ Mpc used in identifying clusters of galaxies.

The smoothing length $R_b$ introduces another parameter to our model. To study the influence of the smoothing length to the supercluster definition we test three values: $R_b = 8$, 16 and 24 $h^{-1}$ Mpc. In Figure 2 we plot equidensity contours for all three smoothing lengths. These calculations show that the mean density contours (the lowest contour level plotted) are almost independent on the smoothing parameter $R_b$, provided $R_b \gg R_g$. In the following we use smoothing length $R_b = 8\ h^{-1}$ Mpc to identify superclusters. To avoid merging of nearby superclusters we use a density threshold $D_t = 1.1$, higher than the threshold to divide the matter into the clustered and non-clustered components.

Another possibility to define superclusters is selecting them from the cluster distribution using the cluster analysis as used by EEDTA to investigate the large-scale distribution of clusters and superclusters. For the present analysis we prefer the use of the smoothed density field. In 3-dimensional models we have used both methods to define superclusters, clusters located in rich superclusters with at least 4 rich clusters as members are plotted in Figure 9 by filled circles.

## 3. STATISTICAL TESTS

### 3.1. Mass distribution of clusters of galaxies

In order to compare our cluster catalogue with observations we have calculated the cluster mass function. The observed cluster mass function, found by Bahcall and Cen (1993), is given as the integrated function

$$N(M) = \int_M^\infty n(m)dm, \qquad (6)$$

where $n(m)dm$ is the number of clusters in the mass interval $m$ to $m + dm$. The mass function can be represented by the following Schechter-type analytic expression

$$N(M) = a \left(\frac{M}{M^*}\right)^{-\gamma} \exp(-M/M^*). \qquad (7)$$

**Fig. 2.** Equidensity contours of the model M2p with density smoothing at 8, 16, and 24 $h^{-1}$ Mpc smoothing length (from left to right). Note the approximate similarity of the lowest contour in all three cases.

**Table 2.** Parameters of the Mass Distribution Function

| Model | $a$ | $M^*$ | $\gamma$ |
|---|---|---|---|
| | $10^{-5} h^{-3}$ Mpc$^3$ | $10^{15}\ M_\odot$ | |
| M2pk2 | 9.4 | 0.13 | $-0.53$ |
| M2pk4 | 0.7 | 0.37 | $-1.32$ |
| M2pk8 | 0.1 | 0.57 | $-1.75$ |
| M2pi1 | 0.3 | 0.43 | $-1.58$ |
| M2pi2 | 2.1 | 0.27 | $-0.90$ |
| M2pi3 | 1.4 | 0.38 | $-0.83$ |
| HDM | 2.2 | 0.38 | $-0.59$ |
| M1p | 0.6 | 0.37 | $-1.37$ |
| M1pt | 8.3 | 0.12 | $-0.70$ |
| M2p | 1.0 | 0.34 | $-1.23$ |
| N2p | 1.6 | 0.26 | $-1.08$ |
| CDM1 | 13.1 | 0.04 | $-0.50$ |
| CDM2 | 3.8 | 0.23 | $-1.28$ |

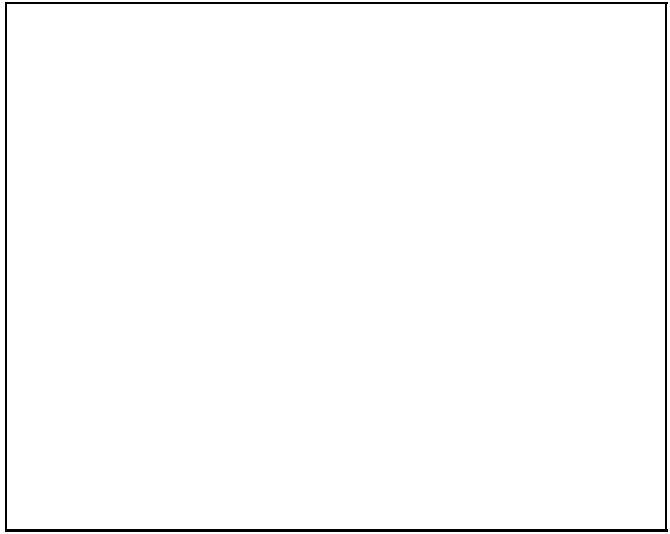

**Fig. 4.** The distribution of supercluster masses. Dots note the observed mass distribution, calculated on the basis of the supercluster catalogue by EETDA.

Parameters $a$, $\gamma$, and $M^*$ are found for all models and are given in Table 2. By comparing models with observations we found parameters $a$ and $M^*$ in physical units.

The cluster mass function for models is plotted in Figure 3 together with the observed function. The Figure shows that our models reproduce the observed mass function surprisingly well. Only in the case of poor clusters of galaxies the number of clusters in models is smaller than given by observations. In the following we shall use basically rich clusters of galaxies. For rich clusters there is no systematic deviation of models from observations for the majority of models.

To characterize different models we also use the distribution of masses of superclusters. Results of this comparison for models M1p, M1pt, and M2p are given in Figure 4. Supercluster masses are calculated as described in §2.4 using the density field smoothed on an 8 $h^{-1}$ Mpc scale. Masses are expressed in units of the mean mass of superclusters for the model M1pt.

A proper estimate of the masses of real superclusters is not obvious. For this we need mass estimates of a large number of superclusters either from the velocity field or by counting the number of galaxies in respective superclusters. Presently there is too little information available for a representative sample of superclusters. As a first approxi-mation we can estimate the mass of superclusters from the number of rich clusters present in them. These data can be extracted from the recent catalogue of superclusters by EETDA. The distribution of supercluster masses, again in units of the mean masses of superclusters, is given in Figure 4.

**Fig. 3.** Cluster mass distribution function. Dots represent the observed function according to Bahcall and Cen (1993), lines various models (see the Figure). Note the similarity of mass distribution for models with different large-scale power (upper left panel), and with variable scale length of the maximum (lower left panel).

### 3.2. Correlation function and power spectrum

In Figure 1 we plot the power spectra for all models for the present epoch. In the panel for 3-dimensional models we give also the observed power spectrum according to Gramann and Einasto (1992) and Einasto *et al.* (1993); other recent determinations by Peacock and West (1992), Fisher *et al.* (1993), Park *et al.* (1994) and Peacock and Dodds (1994) give similar results.

We have calculated the mass and cluster correlation functions for all of our models. Clusters are divided into two richness classes using mass limits as defined by the mass distribution function. The correlation length $r_0$, defined as the distance for which the correlation function has the value $\xi(r_0) = 1$, was calculated from the mass correlation function and is given in Table 3.

**Table 3.** Parameters of the Correlation Function

| Model | $r_0$ | $r_{max}$ |
|---|---|---|
| M2pk2 | 5.4 | 175 |
| M2pk4 | 4.9 | 150 |
| M2pk8 | 4.6 | 75 |
| M2pi1 | 4.0 | 120 |
| M2pi2 | 5.5 | 125 |
| M2pi3 | 9.0 | 150 |
| HDM | 19.2 | 130 |
| M1p | 5.5 | – |
| M1pt | 5.2 | 120 |
| M2p | 5.0 | 120 |
| N2p | 5.5 | 150 |
| CDM1 | 3.7 | – |
| CDM2 | 5.1 | – |

**Fig. 5.** The correlation function. Dots are for the observed correlation function for rich clusters of galaxies (Einasto *et al.* 1992) for the 300 $h^{-1}$ Mpc sample in the Northern supergalactic hemisphere. For models the correlation function of poor clusters of galaxies is given.

The cluster correlation function has a weak secondary maximum at $\approx 130\ h^{-1}$ Mpc, as found for the Northern supergalactic cluster sample by Einasto *et al.* (1992). Most cluster correlation functions of our simulated samples also have a weak secondary maximum at $r_{max}$. This value is given in Table 3; the correlation function for simulated clusters for all model samples is given in Figure 5. The observed correlation function for Abell clusters is taken from Einasto *et al.* (1992).

To understand better the geometrical origin of the secondary maximum we constructed, following Einasto (1992), a series of "toy models". The analysis of toy models will be published elsewhere, here we summarize only the main results of this study. Toy models have a built-in scale $\approx 130\ h^{-1}$ Mpc, either in the form of a rectangular grid or as a mean separation between voids. A sharp secondary maximum is observed in all models with a fixed grid, the strength of the maximum depends on the distribution of clusters, whether they are located at grid corners, edges or surfaces. The maximum is shallower if the grid size is changing within certain limits around the mean value. The secondary maximum is absent in the case of randomly located voids.

These toy models demonstrate that the secondary maximum is due to the presence of a quasi-regular pattern in the distribution of clusters. The observed secondary maximum in the cluster correlation function is not very strong, thus the regularity is not pronounced. The mass correlation function has no secondary maximum, exceptions are only models M2pi3 and HDM, where both the mass and cluster correlation functions have a secondary maximum at the same location.

### 3.3. Void diameter statistics

One kind of statistics we use in this paper is the mean diameter distribution of voids defined by different objects: rich and poor clusters of galaxies and single galaxies. In the case of galaxies all test particles above the threshold density ($D_t = 1$) were considered as galaxies. Rich and poor clusters were defined as described above. Rich clusters correspond to Abell clusters, poor clusters to nearby Zwicky clusters.

Voids were found using the empty sphere method by Einasto et al. (1989): the volume under study was divided into small cubic cells of size $l = L/k$, where $L$ is the size of the whole volume under study, and $k$ is a resolution parameter. Then for each cell center the distance to the nearest galaxy in the whole sample was found. Cells where this distance has a maximum are located in centers of voids, and the respective distance corresponds to the void radius. The void search algorithm finds any maxima in the distance field. We want to study well-defined voids, i.e. voids surrounded from all sides by galaxies. To avoid ill-defined voids we exclude from the void sample all voids with center coordinates close to sample boundaries.

Most voids are elongated, and our program identifies several close voids with almost identical void radius. Lindner et al. (1994) studied the effect of overlapping voids and demonstrated that this phenomenon has little effect on mean void diameter statistics. Thus we can ignore this effect.

A technical parameter in void search is the resolution parameter $k$. Lindner et al. (1994) studied the influence of this parameter and showed that statistically the mean void diameter is independent of the value of this parameter, however, the higher this parameter the smaller are random fluctuations of results. For this reason we have used high values, $k = 256$ and $k = 32$ in 2- and 3-dimensional cases, respectively.

The size of voids defined by clusters depends critically on the number density of clusters. In the 3-dimensional case we use the number density of rich clusters (Abell and ACO clusters of all richness classes) found by Bahcall and Cen (1993), $13.5 \times 10^{-6}$ $h^3$ Mpc$^{-3}$; the density of Zwicky clusters is 5.6 times higher.

Mean diameters of voids for a Poisson distribution of particles are for simulated galaxies in our 3-dimensional simulations 7.4 $h^{-1}$ Mpc, for simulated Abell and Zwicky clusters 78 $h^{-1}$ Mpc and 44 $h^{-1}$ Mpc, respectively. Actual mean diameters of voids are larger, thus in the 3-dimensional case the void statistics is meaningful.

In our 2-dimensional calculations the mean diameter of galaxy-defined voids for a Poisson distribution is 1.7 $h^{-1}$ Mpc, that of rich cluster-defined voids about 100 $h^{-1}$ Mpc, approximately equal to the mean size of voids in the observed cluster distribution (EEDTA). Thus the number density of clusters in 2-dimensional simulations is too low to apply the void statistics. For this reason we give in Table 4 mean diameters of cluster-defined voids only for 3-dimensional models.

**Table 4.** Mean diameters of cluster defined voids

| Model | $D_{rich}$ | $D_{poor}$ |
| --- | --- | --- |
| N2p | $91.3 \pm 11.6$ | $51.3 \pm 10.1$ |
| CDM1 | $82.6 \pm 12.5$ | $47.6 \pm 7.2$ |
| CDM2 | $90.5 \pm 14.4$ | $47.5 \pm 9.8$ |

Diameters of galaxy-defined voids were calculated for a series of 2-dimensional simulations for three different epochs, separately for high- and low-density regions (inside and outside of superclusters). Results are given in Table 5. We give also the scatter of diameters, this number is not to be confused with the statistical error of void diameters.

## 4. THE INFLUENCE OF THE POSITION OF THE MAXIMUM OF THE SPECTRUM

Models M2pk2 through M2pk8 can be interpreted in two different ways. One interpretation is that the physical size of these three models is the same, say $\approx 500$ $h^{-1}$ Mpc, and the maximum of the spectrum, $\lambda_t$ has three different values: $\approx 65, 130$ $h^{-1}$ Mpc and $\approx 260$ $h^{-1}$ Mpc. The second interpretation is that $\lambda_t \approx 130$ $h^{-1}$ Mpc but the size of the box, $L$, changes from $\approx 250$ $h^{-1}$ Mpc to $\approx 1000$ $h^{-1}$ Mpc. In this Section, we use the former interpretation in order to investigate whether properties of the large-scale structure can be changed simply by tuning the length scale of the initial power spectrum.

In upper panels of Figure 6 we give the distribution of particles representing galaxies, in lower panels the distribution of the smoothed (at $R_b = 8$ $h^{-1}$ Mpc smoothing length) density field. We see clearly the concentration of simulated galaxies and superclusters into circular structures which are similar to structures observed by Vettolani et al. (1994a, 1994b) in a deep 2-dimensional survey of faint galaxy redshifts. The scale of these structures in the models varies: in model M2pk2 the circular structures have a diameter about half of the computational box, in model M2pk4 the structures are about two times smaller, and in model M2pk8 they are smaller by another factor of two.

Now we consider the cluster correlation function. The secondary peak of the correlation function characterizes the mean distance between neighboring superclusters. In Figure 5 for these models the correlation function is plotted under the assumption that the size of the computational box for all models is 512 $h^{-1}$ Mpc. We see that in model M2pk2 the secondary peak of the correlation function is located at $\approx 175$ $h^{-1}$ Mpc. In model M2pk4 we see a maximum at $\approx 150$ $h^{-1}$ Mpc, in model M2pk8 – at distance $\approx 75$ $h^{-1}$ Mpc. A look at Figure 6 confirms that these maxima correspond to supercluster distances across voids.

The observed mean distance between superclusters across voids is $\approx 130$ $h^{-1}$ Mpc, regardless of the position of the region on the sky. In other words, this scale-length is an invariant observational parameter. This means that the position of the maximum in the spectrum is not a free parameter. Because it defines the scale of the cellular structure of the large-scale distribution of clusters and galaxies, it is fixed by the observations.

**Fig. 6.** The distribution of simulated galaxies and the density field (smoothed with dispersion 8 $h^{-1}$ Mpc ) in models M2pk2, M2pk4 and M2pk8.

## 5. INFLUENCE OF SMALL-SCALE POWER

To investigate the influence of the change of the amplitude of small-scale fluctuations we use models M2pi1, M2pi2, M2pi3 and HDM. In these models the initial amplitude of small-scale perturbations with respect to medium-scaled perturbations decreases when we move to larger negative power index, and is absent in the HDM model.

The basic parameter which varies in this series of models is the number of massive clusters relative to the number of low-mass clusters. This change can be characterized by the slope $\gamma$ of the mass function (see Table 2 and Figure 3). Model M2pi1 agrees best with observations. In other models the slope of the mass function is too small, i.e. there are too many very rich clusters in the model. The HDM model deviates most from the observations. This difference in mean masses of clusters can be seen in the distribution of particles shown in Figure 7.

Another aspect of small-scale fluctuations is the presence of galaxies in cluster-defined voids (supervoids). If initially there is no power on small scales then in supervoids no galaxies form at all. Our HDM model corresponds just to this case.

During the evolution of the HDM model also small scale fluctuations form, and at the present epoch the slope of the spectrum in the HDM model on small scales is $m_3 \approx -3$, close to the slope $-3$ in model M2pi3 (see Figure 10 and further discussion of the evolution in §8). The structure of these models is, however, very different. In model M2pi3 (middle panel of Figure 7) galaxy filaments are located over the *whole space* including supervoids, while in the *HDM model there are no galaxies in supervoids*. In other words, small scale power which forms in the HDM model during the evolution, is generated by structures located in high-density regions, low-density regions remain empty.

This difference in the evolution can be expressed also by the statistics of galaxy-defined voids. Mean sizes of galaxy-defined voids are given in Table 5, separately for high- and low-density regions (within superclusters and supervoids). This table demonstrates that in the HDM model galaxy-defined voids in low-density regions have the same size as cluster-defined voids in other models. This result is also expected: visual inspection of the distribution of simulated galaxies shows a complete absence of galaxies and clusters of galaxies in supervoids. In high-density

**Fig. 7.** The distribution of simulated galaxies and the density field in models M2pi1, M2pi3 and HDM.

**Table 5.** Mean diameters of galaxy defined voids

| Model  | $D_{low}$      | $D_{high}$    | $F_{low}$ |
|--------|----------------|---------------|-----------|
| M1p.1  | $10.2 \pm 3.0$ | $7.3 \pm 2.1$ | 0.556     |
| M1p.2  | $12.9 \pm 4.1$ | $8.6 \pm 2.8$ | 0.599     |
| M1p.4  | $20.9 \pm 8.0$ | $11.6 \pm 4.4$| 0.711     |
| M1pt.1 | $10.2 \pm 3.1$ | $7.4 \pm 2.2$ | 0.530     |
| M1pt.2 | $13.0 \pm 4.3$ | $8.8 \pm 2.7$ | 0.565     |
| M1pt.4 | $22.0 \pm 8.0$ | $12.1 \pm 4.0$| 0.664     |
| M2p.1  | $9.6 \pm 2.8$  | $6.9 \pm 2.0$ | 0.527     |
| M2p.2  | $13.0 \pm 4.4$ | $8.9 \pm 2.6$ | 0.566     |
| M2p.4  | $21.4 \pm 7.1$ | $12.4 \pm 3.9$| 0.655     |
| HDM.1  | $92.7 \pm 29.5$| $4.3 \pm 2.3$ | 0.468     |
| HDM.2  | $92.7 \pm 35.0$| $4.2 \pm 2.0$ | 0.573     |
| HDM.4  | $104.8 \pm 44.5$| $3.8 \pm 3.5$| 0.725     |

regions galaxy-defined voids are much smaller than in all other models.

This analysis shows that filamentary structure of galaxies in low-density regions develops only if there is some power on small scales *initially*. It is not sufficient to compare spectra at the present epoch, also the evolutionary history of the structure is important.

## 6. INFLUENCE OF LARGE-SCALE POWER

To investigate the influence on the large-scale power to the formation of the structure we use models M1p, M2p and M1pt. These models differ by the strength of the large-scale power. In model M1p the spectrum on large scales has the same index as on small scales, in this model long wavelength perturbations have the largest amplitude. In model M2p the large-scale power spectrum has the Harrison-Zel'dovich index $n_3 = 1$; in model M1pt the amplitude of the spectrum on large scales is zero.

The parameter most influenced by the change in the large scale power is the mass of superclusters. Superclusters were defined using a smoothing length $R_b = 8 \ h^{-1}$ Mpc and a density threshold $D_t = 1.1$ (§2.5). The distribution of supercluster masses is shown in Figure 4, for all three models considered. Masses are expressed in units of mean masses of superclusters for model M1pt. We see that the distribution of supercluster masses is rather different. The truncated model M1pt has superclusters of approximately equal masses, the dispersion of masses (expressed in units of mean masses of superclusters in model M1pt) is only 0.07. In model M2p some superclusters have masses, exceeding mean masses by a factor of six. In model M1p largest superclusters have masses over ten times the mean mass.

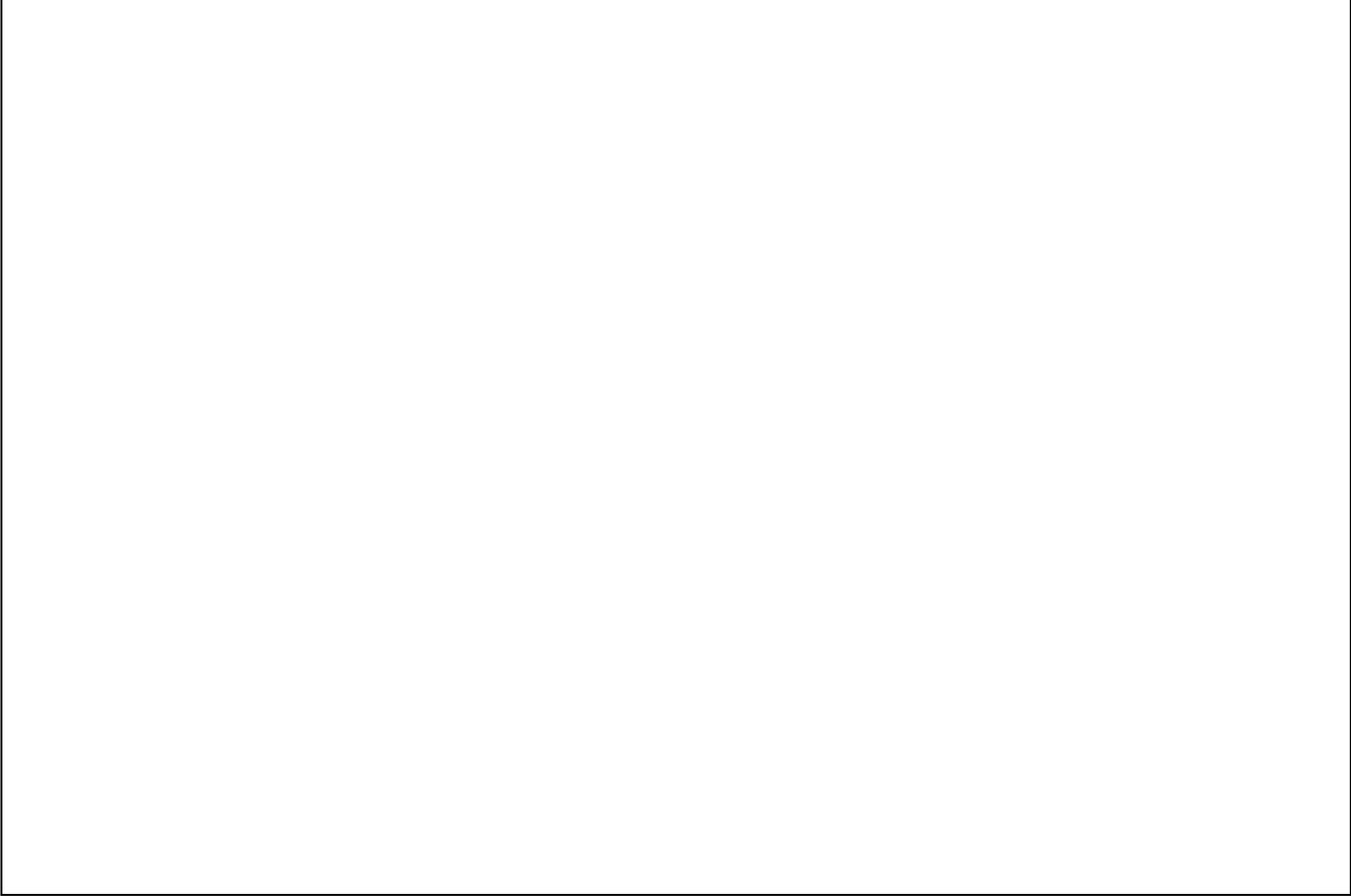

**Fig. 8.** The distribution of simulated galaxies and the density field in models M1p, M1pt and M2p.

The observed mass distribution in superclusters (using the number of rich clusters as an indicator of the supercluster mass) is shown in Figure 4 by filled circles. Here we use the data from EEDTA. Although the methods to determine supercluster masses (see §3.1.) were different the general trend of the mass distribution is similar to the trend we see in the models. The model M2p agrees best with observations. In model M1pt there are too few superclusters with large mass, in model M1p there are too many.

The distribution of simulated galaxies and density contours for all three models is shown in Figure 8. Visual inspection of the distribution suggests that superclusters in the model M1pt have almost equal masses. In the model M2p the dispersion is larger and in the model M1p the largest.

Since we have used identical random numbers to generate the initial spectrum on small scales the filamentary structure is similar in all models. What differs is the strength of filaments. A close inspection of plots shows that some filaments which are located in the M1pt model in high-density regions, are in models M2p and M1p situated in low-density regions. This difference is caused by a change in the power of large-scale modes. This demonstrates that large wavelength perturbations modulate the strength of galaxy systems formed by small-scale perturbations.

Void statistics shows that voids defined by galaxies have almost identical size for all three models (see Table 5). This result demonstrates that galaxy-defined voids are insensitive to the large-scale perturbations. Similarly, a look to Figure 3 shows that large-scale modes do not influence the mass distribution of clusters.

To summarize our analysis we can say that most quantitative tests are insensitive to changes in the large-scale modes of density perturbations. The mass of superclusters is the only parameter which depends primarily on these modes.

**Fig. 9.** The distribution of simulated galaxies (upper panels) and clusters of galaxies (lower panels) in models N2p, CDM1 and CDM2. A 64 $h^{-1}$ Mpc thick sheet of 3-dimensional simulations has been plotted. Clusters located in rich superclusters with at least four member clusters are plotted as filled circles, isolated clusters and clusters in poor superclusters as dots.

## 7. INFLUENCE OF THE HEIGHT OF THE MAXIMUM OF THE SPECTRUM

To study the influence of the height of the maximum of the spectrum we use two CDM models and a double power law model with a sharp transition of the power index, eqn. (4).

The distribution of simulated galaxies in both CDM models and the N2p model is shown in Figure 9. Slices 64 $h^{-1}$ Mpc thick are plotted. The presence of high- and low-density regions is well seen in all three models. Since a rather thick sheet is shown and the resolution of the calculations is four times lower than in the 2-dimensional models, the distribution of particles is much smoother than those plotted in Figure 8.

In lower panels of Figure 9 we show the distribution of rich clusters of galaxies in these three models. Clusters located in rich superclusters are plotted as filled circles. The distribution of galaxies and rich clusters of galaxies is rather irregular in model CDM1. In this model there are only a few very rich superclusters as measured by the number of rich clusters in superclusters, and their mutual distance is much larger than in the real Universe. In model CDM2 superclusters are distributed more regularly, and the number of rich superclusters is comparable with the number of real rich superclusters. However, there is no preferred scale of the supercluster-void network. Only in model N2p we see a regular pattern with constant scale.

This visual impression is confirmed by the cluster correlation functions which has no real minimum and secondary maximum in model CDM1, a weak minimum without a secondary maximum in model CDM2, and a well defined minimum and shallow secondary maximum in model N2p.

The cluster correlation function is the basic quantitative statistics which is sensitive to differences in the large-scale distribution of superclusters observed in models CDM1, CDM2 and N2p. The mean diameter of cluster-defined voids in model CDM1 is considerably smaller than in the other two models and is practically equal to the diameter of voids in a Poisson sample with the same number density of objects. This test also demonstrates the difference between the model CDM1 on the one side and models CDM2 and N2p on the other. Other quantitative tests

(cluster mass function, mass correlation function) show no significant differences between these three models.

The absence of a clear pattern in the distribution of superclusters is a generic property of all models with a shallow maximum as it is the case with CDM models. The model CDM2 has more power on large scales but the maximum of the spectrum is not well defined, thus the distribution of rich clusters is organized not so well as in models with a sharp maximum (model N2p).

## 8. EVOLUTION OF THE SUPERCLUSTER-VOID NETWORK

To investigate the evolution of the structure we have stored the distribution of model particles at four epochs, corresponding to density dispersions 0.5, 1, 2, and 4 on the 1 $h^{-1}$ Mpc scale. In order to illustrate the evolution of the spectrum we plot in Figure 10 spectra of these epochs for two models, M2p and HDM. The evolution of the spectrum for the rest of the models is very similar to the evolution of the model M2p. We see that spectra at different epochs of the model M2p run almost parallel to each other, i.e. the evolution is basically linear. In the HDM model the growth of the spectrum on small scales is much more rapid, here a power-law spectrum with index $m_2 = -2$ forms, this corresponds to index $m_3 = -3$ in the three-dimensional case.

The distribution of simulated galaxies in model M2p for three epochs, corresponding to density dispersion 1, 2, and 4, is shown in Figure 11. We plot in this Figure also the density field smoothed on $R_b = 8\ h^{-1}$ Mpc scale, for all three epochs.

Mean void diameters for simulated galaxies for these epochs are given for several models in Table 5. Void diameters were found separately for voids in superclusters ($\rho_8 \geq 1$) and in supervoid regions ($\rho_8 < 1$).

Our results show that in all models voids between filaments in supercluster regions are smaller than in low-density regions. The diameter of voids increases with time. This increase is due to two different processes. First, during the evolution smaller voids join to form larger ones (compare the distribution of particles at epochs 1 and 4 in Figure 11), and second, the dimensions of walls between voids shrink. The difference of void diameters in high- and low-density regions increases with time – in other words, voids in low-density regions expand faster than in high-density regions.

The particle distribution in Figure 11 shows that at the epoch $\sigma = 1$ clusters of galaxies start to form. This epoch can be identified with the beginning of the intensive galaxy formation. Most of the matter in this epoch is still distributed rather smoothly. In the second epoch the clusters are better seen, but only in the third epoch which corresponds to the present time we see a well developed network of clusters, filaments, and superclusters.

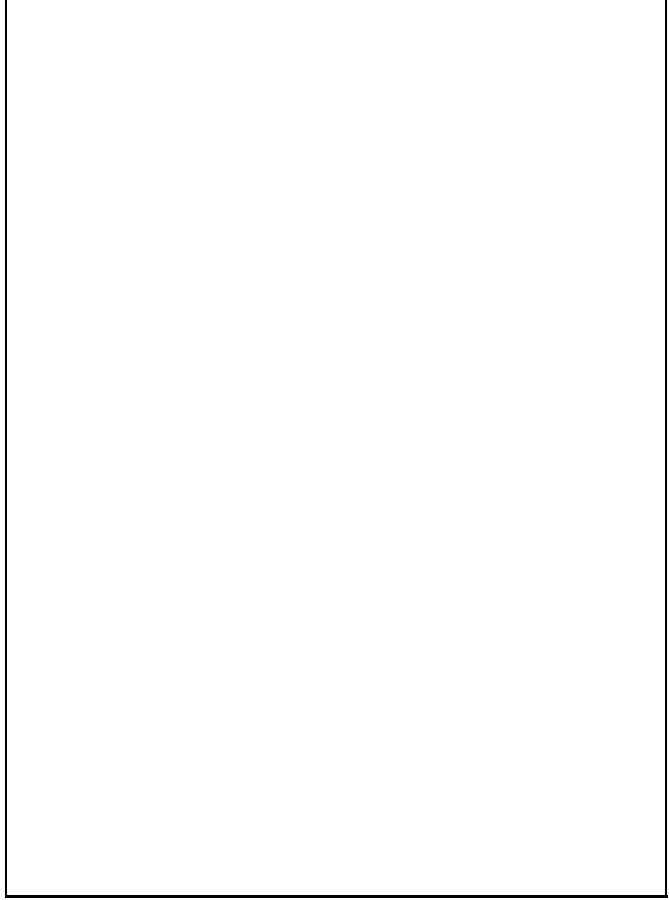

**Fig. 10.** The evolution of spectra of models M2p and HDM. The solid curve denotes initial spectrum, dotted and dashed lines spectra at time steps 1, 2, and 4, as measured by the density dispersion.

This comparison shows that clusters of galaxies are still in the process of being formed, while their richness grows steadily.

The comparison of the smoothed (with dispersion 8 $h^{-1}$ Mpc) density field shows that all superclusters observed in the present epoch can be followed already in the $\sigma = 1$ epoch, only the density contrast is much smaller. All high-density regions can be easily identified at all epochs. During the evolution the density contrast increases, but dimensions of superclusters are rather stable. We see only a very small contraction of filaments and expansion of voids.

## 9. DISCUSSION

### 9.1. Influence of the maximum of the spectrum

A quasi-regular network of superclusters is possible only if there is a well-defined maximum of the power spectrum. Our models M2pk4 and N2p with a sharp transition of the power index probably exaggerate the regularity of the structure, (cf. the distribution of clusters in rich superclusters in these models, Figures 6 and 9, and in observations, Figure 5 by EETDA). On the other hand, if the maximum of the spectrum is very shallow, as in CDM-models, then the distribution of superclusters is rather irregular. The best agreement with observations has probably the model with a smooth but rapid transition between positive and negative spectral index approximated by formula (3).

**Fig. 11.** The evolution of the distribution of simulated galaxies and smoothed density field (smoothing length 8 $h^{-1}$ Mpc). The lowest density contour corresponds to mean density level.

The comparison of models with different scale of the maximum wavelength demonstrates clearly that the mutual distance between superclusters and the scale of the supercluster-void network is determined by the scale of the maximum of the spectrum.

An important problem is the determination of the position of the maximum of the spectrum from observations. Direct determinations of the spectrum on large scales have large random errors since the size of observational samples is close to the scale of interest. In principle, deep one- and two-dimensional samples can give better results. However, not all one- or two-dimensional samples cross supervoids through central regions, thus in most cases only little information on the actual power spectrum can be extracted, see Kaiser and Peacock (1991) for the discussion of this problem.

One possibility to derive the scale of the maximum of the power spectrum is to use data on the correlation function of rich clusters of galaxies on large scales. The scale of the secondary maximum is related to the scale of the maximum of the spectrum. However, a direct relation between these scales is valid only in analytic models (Einasto and Gramann 1993). Our simulations and the study of various toy models has shown that in realistic situations this relationship is only approximate. In many simulations the secondary maximum of the cluster correlation function is very shallow, and the scale of the secondary maximum does not coincide exactly with the scale of the maximum in the power spectrum.

We believe that a combined approach which takes into account all available direct and indirect data is probably the best in the determination of the scale $\lambda_t$. Best available estimates for the scale of the maximum come from the deep pencil-beam survey by Broadhurst *et al.* (1990), from the deep two-dimensional survey by Vettolani *et al.* (1994a,b), from the distribution of superclusters and voids by EEDTA, from the secondary maxi-mum of the correlation function of rich clusters of galaxies by Einasto and Gramann (1993), and from the study of the second derivative of the correlation function by Mo *et al.* (1992a, 1992b). Based on these determinations we adopted $\lambda_t = 128\ h^{-1}$ Mpc. The formal error of this parameter as estimated from the scatter of individual determinations is rather small ($\approx 3\ h^{-1}$ Mpc). The actual error due to cosmic scatter and possible systematic errors is much larger. A new determination of this parameter is an important tasks of the observational cosmology.

*9.2. Influence of large-scale density perturbations*

Simple power-law models are essentially scale-free. Properties of these models have been studied by Efstathiou *et al.* (1988), Weinberg and Gunn (1990) and others. Our double power-law models give us the means to investigate the influence of large-scale perturbations to scaling properties of galaxies.

Data discussed in earlier Sections demonstrate that most quantitative statistics are insensitive to large-scale perturbations. The mass correlation function, mass distribution of clusters, and the distribution of galaxy-defined voids is essentially determined by medium and small-scale perturbations, and are practically not influenced by large-scale perturbations. These statistics define directly or indirectly a certain scale (correlation length, mean void diameter etc), and this scale is practically the same for all models, provided small-scale perturbations as measured by the power spectrum on medium and small scales are identical.

The insensitivity of scaling properties to large-scale density perturbations does not mean that these perturbations have no influence at all. Our analysis has demonstrated that large-scale modes determine the geometric pattern of the supercluster-void network, supercluster mass distribution, and the strength of filaments in different environment. These aspects of the large-scale distribution are more difficult to express in quantitative terms. The geometric pattern of superclusters and voids can be measured by the sizes of voids, and by the cluster correlation function on large scales. The mass distribution function of superclusters can be estimated in models directly by counting the mass in all high-density regions and indirectly by the number of rich clusters in superclusters. In the real world only this indirect approach can be applied presently to determine the mass of superclusters. The supercluster mass distribution test gives preference to models with a Harrison-Zel'dovich spectral index on large scales.

*9.3. Fine structure of supervoids*

The basic result of the study by Lindner *et al.* (1994) was the establishing of the fine structure and void hierarchy in low-density regions (supervoids). In that study we investigated the Northern Local Void, but plots of galaxies in other regions of the sky demonstrate that this is a common property of supervoids.

Our study shows that the presence of some initial power on small and intermediate scales in the models is crucial for the development of fine structure in supervoids. Models with some power on small scales can be called CDM-type. The fine structure is absent in all HDM-type models where there is no power on small scales initially. The formation of some power on small scales in later epochs does not change the situation – our calculations show clearly that fine structure forms only in high-density regions of HDM models.

Another manifestation of the difference between CDM- and HDM-type models is the void diameter statistics. Galaxy-defined voids in CDM-type models are considerably smaller than cluster-defined voids. In HDM-type models galaxy-defined voids are identical in diameter to cluster-defined voids in the real Universe and in most of our simulations. In other words, there is no void hierarchy in HDM-type models.

This difference between CDM-type and HDM-type models may be called in terms of different structure formation concepts – "top-down" and "bottom-up". In the classical top-down scenario first objects to form are superclusters of galaxies which during the evolution fragment into galaxies and small systems of galaxies. In the classical bottom-up scenario the formation starts from small units which merge during the evolution to form systems of larger scale. In particular, superclusters form in the second scenario only gradually in later epochs, and their growth should continue indefinitely.

Our calculations show that either picture is too simplified. The location of superclusters is given already by the initial conditions: the large-scale perturbations determine the location of superclusters and supervoids. Furthermore, clusters and filaments of galaxies are influenced by the large-scale environment: in supervoids these systems remain poor, in superclusters they are richer from the very beginning of the structure formation. A similar conclusion has been drawn also by Kofman *et al.* (1992) on the basis of two-dimensional adhesion model of the structure evolution.

On the other hand, the formation of fine structure in the top-down scenario is not sufficient to produce filaments in supervoids. The presence of small-scale structure in the Universe was first demonstrated using the multiplicity function test (Zeldovich, Einasto and Shandarin 1982, Einasto *et al.* 1984). The HDM-model lacks this property, there are only systems of large multiplicity. The present study demonstrates that this difference is more fundamental. It is not sufficient to produce the filamentary structure in superclusters only, the HDM-type models can, in fact, achieve this in later evolutionary stages, but such structure must also be formed in supervoids, where the HDM-type models fail completely.

Recently, interest in the HDM-model has been revived (Beaky *et al.* 1992, Blanchard *et al.* 1993). A number of models were presented and it has been suggested that these models can predict the observed structure. According to our results, these suggestions obviously ignore the presence of filamentary systems in low-density regions, which cannot be accounted for in HDM-models.

To summarize, the comparison of models with observations shows that conventional HDM and CDM models are excluded by the combined evidence presented here. The best agreement with observations was found for a model with a Harrison-Zel'dovich spectrum on large scales, a CDM-like spectrum on small scales, and with a rapid transition from positive to negative power index. Such spectrum may be considered as a combination of HDM and CDM models (Davis, Summers and Schlegel 1992, Holtzman and Primack 1993, Klypin *et al.* 1993), or interpreted as well in terms of the primeval baryon isocurvature model (Peebles 1987).

## 10. CONCLUSIONS

In this paper we have investigated the influence of perturbations of different scale to the large-scale properties of galaxies and clusters. The principal results of this study are:

1) a quasi-regular network of superclusters and voids forms only in models with a well-defined maximum of the spectrum. The scale of the supercluster-void network is given by the scale of the maximum of the spectrum; presently available estimates suggest $128\ h^{-1}$ Mpc;
2) the fine structure in large low-density regions (supervoids) forms in all models where small-scale perturbations are present *initially* and is absent in HDM-type models where small-scale perturbations form only later;
3) the mass distribution of clusters of galaxies is determined by small-scale and intermediate perturbations;
4) large-scale perturbations (wavelength larger than the scale of the maximum of the spectrum) modulate the strength of the structures, determined by small-scale and intermediate perturbations;
5) supercluster-supervoid structure forms in a very early stage of the evolution of the Universe from large-scale density fluctuations; the sizes of voids increase with time due to merging of voids. Voids between filaments in superclusters are considerably smaller than voids in the low-density environment;
6) present tests exclude the conventional HDM and CDM models. The model which agreed best with the observations is the double power model with a Harrison-Zel'dovich spectrum on large scales and an index $n_3 \approx -1.5$ on small scales.

*Acknowledgements*. This study started during a visit of JE and ME to European Southern Observatory, and continued during visits to Göttingen University Observatory. JE and ME thank the staff of these Observatories for support and very stimulating atmosphere, allowing to establish this collaboration. Fruitful discussions with Prof. N. Bahcall and Drs. E. Saar and D. Weinberg are acknowledged. This study was supported by Estonian Science Foundation grant 182 and International Science Foundation grant LDC 000. PF was partly funded by the "Studienstiftung des Deutschen Volkes".